\documentclass{iopart}
\usepackage{iopams}
\usepackage{graphicx}
\usepackage{setstack}

\def\eps{\boldsymbol {\epsilon}}
\def\fh{\hat{f}}
\def\gh{\hat{g}}

\newcommand{\enc}{l}

\begin{document}

\title{Moments of the {W}igner delay times}

\author{Gregory Berkolaiko$^1$ and Jack Kuipers$^2$}
\address{$^1$ Department of Mathematics, Texas A\&M University, College Station, TX 77843-3368, USA}
\address{$^2$ Institut f\"ur Theoretische Physik, Universit\"at Regensburg, D-93040 Regensburg, Germany}
\ead{Jack.Kuipers@physik.uni-regensburg.de}

\begin{abstract}
The Wigner time delay is a measure of the time spent by a particle inside the
scattering region of an open system.  For chaotic systems, the
statistics of the individual delay times (whose average is the Wigner time delay)
are thought to be well described by random matrix theory.  Here we
present a semiclassical derivation showing the validity of random matrix results.
In order to simplify the semiclassical treatment, we express the moments of the delay times in terms of
correlation functions of scattering matrices at different energies.  In the
semiclassical approximation, the elements of the scattering matrix
are given in terms of the classical scattering trajectories, requiring 
one to study correlations between sets of such trajectories.
We describe the structure of correlated sets of trajectories and
formulate the rules for their evaluation to the leading order in
inverse channel number.  This allows us to derive a polynomial
equation satisfied by the generating function of the moments.  Along
with showing the agreement of our semiclassical results with the
moments predicted by random matrix theory, we infer that the
scattering matrix is unitary to all orders in the semiclassical
approximation.
\end{abstract}

\pacs{03.65.Sq, 05.45.Mt}

\section{Introduction} \label{introduction}

The dynamics of an open quantum system can be described by its
scattering matrix.  The scattering matrix is defined as the linear
operator which transforms an incoming wavefunction, expanded in an
asymptotic channel eigenmode basis, into the outgoing wavefunction.  
Probability conservation forces the scattering matrix to be
unitary.  As the scattering matrix describes the system, it
can be used to investigate the desired physical properties of the
system.

A quantity of particular interest is the Wigner time delay
\cite{eisenbud48, wigner55} which was first derived for the one
channel case from a Hermitian operator based on the scattering
amplitude.  The time delay is, as the name suggests, a measure of the
extra time a particle spends inside the scattering region as a result
of being scattered (for some details see \cite{aj07} for example), 
and was later generalised to multi-channel
scattering matrices \cite{smith60}.  With $M$ scattering channels, the
Hermitian operator admits $M$ eigenvalues which are the {\em
  individual delay times\/} of the system, and the {\em Wigner time
  delay\/} is simply the average of these values.

For classically chaotic systems with a small opening, the probability
for particles to remain inside the system decays exponentially with a
typical time scale depending on the size of the opening.  The
exponential decay can be seen as a natural consequence of the
ergodicity of the classical motion as (for reasonable times) the
particle is equally likely to hit any part of the boundary of the
system leading to a roughly constant small probability to leave
through the opening each time it hits.  The continuous limit of this
process is the exponential decay, and the time scale associated to the
decay is, for chaotic systems, exactly the average time delay
\cite{lv04b} of the corresponding quantum system.  Staying with the
classical version for now, if we spread particles evenly over the
available space and evolve them over time they will condense
onto a typically fractal pattern arranged around the zero-measure set
of trapped periodic orbits that never leave the system.  From this
set, and following their stable manifolds very closely we can
construct trajectories that start outside the system, approach trapped
orbits and remain inside the system for arbitrarily long times before
eventually escaping following the unstable manifolds.  In fact, it
turns out that these trajectories, or rather correlations between
them, are responsible in the semiclassical limit for recreating the
oscillating part of the time delay \cite{ks08}, which is given exactly
in terms of the trapped periodic orbits \cite{bb74,val98}.

As the oscillating part of the time delay is given in terms of
periodic orbits in the semiclassical limit (of $\hbar\to0$),
correlations between periodic orbits must then be responsible for the
typical fluctuations of the Wigner time delay.  In particular the form factor
of the time delay can be written in terms of pairs of periodic orbits
and the types of correlations that contribute were first treated for
the spectral form factor for closed systems, where Gutzwiller's trace
formula \cite{gutzwiller71,gutzwiller90} likewise provides a sum over
pairs of orbits.  The semiclassical treatment of this sum started with
the diagonal approximation \cite{berry85} of pairing orbits with
themselves (or their time reversal) and used the sum rule of
\cite{ha84} to find the semiclassical contribution.  This was followed
more recently by the treatment of correlated pairs of orbits which are
almost identical everywhere, but which differ in a very small region
called an encounter where the orbits behave differently and end up
reconnecting inside the encounter \cite{sr01}.  All the possible types
of orbits with encounters were then generated and treated
\cite{mulleretal04,mulleretal05}, and this treatment was applied to
the time delay to obtain a semiclassical expansion for its form factor
\cite{ks07b}.

The expansion for the form factor of the Wigner time delay was shown
to agree with the result from random matrix theory (RMT), in line with
the idea that most properties of quantum chaotic systems are well
described by the results of RMT.  But RMT can tell us at lot more
about the expected typical behaviour of the individual delay times,
and in particular that their distribution should have a compact
support \cite{bfb99} in the limit where the opening supports a large
number of channels.  Of course the classical decay gives no upper
bound on the distribution of the delay times, and so (if we expect to
recover RMT results) quantum interference as expressed through
semiclassical correlations between classical trajectories should
somehow contrive to limit the maximum delay time.

Our aim in this article is to identify the sort of correlations that
contribute in the semiclassical limit and to describe how they lead to
the RMT result.  We start with an introduction to the time delay in
section~\ref{timedelay}.  To study the distribution of the delay times
we evaluate their moments which can be expressed in terms of correlation
functions of the scattering matrix.  This approach simplifies
the semiclassical treatment compared to the direct evaluation of the
time delay matrix.

In section~\ref{rmtpredictions} we briefly review the RMT results
before applying the semiclassical approximation to the scattering
matrix elements to obtain expressions in terms of scattering
trajectories.  For the low order moments, which we treat in
section~\ref{firstmoments}, we build on the work of the semiclassical
treatment of the conductance \cite{rs02,heusleretal06} and its second
moment \cite{braunetal06,mulleretal07}.  To treat all the moments, we
delve into the combinatorial relations that first arose in the
treatment of the moments of the transmission amplitudes \cite{bhn08}.
This work is extended in section~\ref{allmoments}, where we find an
implicit expression for the generating function of the moments and
show complete agreement with the RMT distribution of delay times.
Considering a simpler case in~\ref{sec:C_all_orders}, we briefly derive the correlation
functions which recently appeared in the semiclassical treatment of
the density of states of Andreev billiards \cite{kuipersetal09}.

\section{The time delay matrix} \label{timedelay}

If we consider a chaotic cavity with one or more open leads that carry
$M$ scattering channels, the scattering dynamics is encoded in the $M
\times M$ unitary scattering matrix $S(E)$ which relates the incoming
and outgoing waves.  We are interested in the Wigner time delay, which
represents the extra time spent in the scattering process compared to
free motion, and which can be found using the Wigner-Smith matrix
\cite{eisenbud48, wigner55, smith60}
\begin{equation} 
  \label{eq:Q_def}
  Q = \frac{\hbar}{\rmi} S^{\dagger}(E) \frac{\rmd S(E)}{\rmd E} .
\end{equation}
Differentiating the unitarity condition, $S^{\dagger}(E)S(E)=I$, with
respect to $E$, we see that
\begin{equation} 
  \label{Ssimpeqn1} 
  S^{\dagger}(E)\frac{\rmd S(E)}{\rmd E} 
  = -\frac{\rmd S^{\dagger}(E)}{\rmd E} S(E) , 
\end{equation}
and that the matrix $Q$ is Hermitian with real eigenvalues.  
The $M$ eigenvalues are the individual delay times of the system, 
and the average value of these times is the Wigner time delay
\begin{equation}
  \label{eq:wtd_defn}
  \tau_{\mathrm{W}}(E)  = \frac{1}{M}\Tr \left[Q\right] .
\end{equation}
The moments of the eigenvalues are given by
\begin{equation} 
  \label{timedelayeqn}
  m_{n}=\frac{1}{M}\Tr \left[Q^{n}\right] ,
\end{equation}
and can be used to recover the complete distribution of the
eigenvalues of $Q$.

We can also obtain the Wigner time delay from a
correlation function of the scattering matrix
\begin{equation}
  \label{eq:cepseqn} 
  C(\epsilon) = \frac{1}{M} \Tr\left[
    S^{\dagger}\left(E-\frac{\epsilon\hbar\mu }{2}\right)
    S\left(E+\frac{\epsilon \hbar\mu}{2}\right)
  \right] ,
\end{equation}
where it is convenient to specify the energy difference in units of
$\hbar \mu$, where $\mu$ is the classical escape rate of the
system. This correlation function provides a symmetrized version of
the time delay \cite{lv04a}
\begin{equation} 
  \label{timedelaytrajsym}
  \tau_{\mathrm{W}} = \frac{1}{\rmi\mu}\frac{\rmd}{\rmd
    \epsilon}C(\epsilon)\Big\vert_{\epsilon=0} 
  = \frac{\hbar}{2\rmi M} 
  \Tr\left[S^{\dagger}(E)\frac{\rmd S(E)}{\rmd E}   
    - \frac{\rmd S^{\dagger}(E)}{\rmd E} S(E) \right] ,
\end{equation}
which agrees with the definition in \eref{eq:wtd_defn} because of the
unitarity of the scattering matrix as expressed through \eref{Ssimpeqn1}.

As well as giving us the first moment of the delay times (the Wigner time delay), this
correlation function can also provide us with the second.
Differentiating \eref{Ssimpeqn1} again we obtain
\begin{equation} \label{Ssimpeqn2}
  S^{\dagger}(E)\frac{\rmd^2 S(E)}{\rmd E^2} 
  + \frac{\rmd^2 S^{\dagger}(E)}{\rmd E^2} S(E) 
  =  -2 \frac{\rmd S^{\dagger}(E)}{\rmd E} \frac{\rmd S(E)}{\rmd E} ,
\end{equation}
which can be used to simplify the second derivative of $C(\epsilon)$ to
\begin{equation}
  \frac{\rmd^{2}}{\rmd \epsilon^{2}} C(\epsilon)\Big\vert_{\epsilon=0} 
  = - \frac{\hbar^2\mu^2}{M} \Tr \left[ 
    \frac{\rmd S^{\dagger}(E)}{\rmd E} \frac{\rmd S(E)}{\rmd E}\right] .
\end{equation}
Inserting the identity matrix in the form $S(E)S^{\dagger}(E)$, we obtain
\begin{equation} 
  \label{secondmomenteqn}
  \frac{\rmd^{2}}{\rmd \epsilon^{2}} C(\epsilon)\Big\vert_{\epsilon=0} 
  =  \frac{(\rmi \mu)^2}{M} \Tr\left[Q^{\dagger}Q \right]
  = (\rmi\mu)^{2}m_{2} ,
\end{equation}
and the second moment of the delay times. 

\subsection{Various approaches to higher moments}
\label{highermoments}

Differentiating the correlation function $C(\epsilon)$ further cannot
produce higher moments since the $n$-th moment involves the first
derivative of $S(E)$ appearing $n$ times while the definition of
$C(\epsilon)$, see~\eref{eq:cepseqn}, involves only two matrices.
One way to resolve this difficulty is to define a higher correlation
function
\begin{equation}
  \label{eq:Cvecepsn}
  C(\eps,n) =\frac{1}{M}\Tr\left[ \prod_{j=1}^{n}
    S^{\dagger}\left(E-\frac{\epsilon_{j}\hbar\mu}{2}\right)
    S\left(E+\frac{\epsilon_{j} \hbar\mu }{2}\right)
  \right] ,
\end{equation}
where $\eps=(\epsilon_{1},\ldots,\epsilon_{n})$ is the vector of
energy differences.  From the $n$-th correlation function we can
obtain all the moments up to the $(2n)$-th by differentiating.  The
formulae, which are different for odd and even moments, are
\begin{eqnarray}
  \label{eq:odd_mom}
  m_{2k-1} &= \frac{1}{(\rmi \mu)^{(2k-1)}}
  \left(\prod_{j=1}^{k-1}\frac{\rmd^{2}}{\rmd \epsilon_{j}^{2}}\right)
  \frac{\rmd}{\rmd \epsilon_{k}} C(\eps,n)\Big\vert_{\eps=0} , \\
  \label{eq:even_mom} 
  m_{2k} &= \frac{1}{(\rmi \mu)^{2k}}
  \left(\prod_{j=1}^{k}\frac{\rmd^{2}}{\rmd \epsilon_{j}^{2}}\right)
  C(\eps,n)\Big\vert_{\eps=0} .
\end{eqnarray}

Though this provides an efficient way to calculate the lower order moments, 
the different energy arguments complicate the semiclassical treatment.  
On the other hand, setting all energy
arguments to $\epsilon$ and differentiating $n$ times leads to
appearance of additional unwanted terms.  To deal with such terms we
define the correlation function
\begin{equation} 
  \label{eq:D_def}
  D(\epsilon,n) =\frac{1}{M}\Tr\left[
    S^{\dagger}\left(E-\frac{\epsilon\hbar\mu}{2}\right)
    S\left(E+\frac{\epsilon\hbar\mu}{2}\right)-I\right]^n .
\end{equation}
Because of the relation
\begin{equation}
  \frac{1}{n!}\frac{\rmd^{n}}{\rmd\epsilon^{n}}
  \left[f(\epsilon)-f(0)\right]^{n}\Big\vert_{\epsilon=0} 
  = \left[f'(0)\right]^{n} ,
\end{equation}
we can see that the moments of the time delay matrix are now given directly by
\begin{equation} 
  \label{eq:mom_thru_D}
  m_{n} = \frac{1}{(\rmi\mu)^{n}n!}
  \frac{\rmd^{n}}{\rmd\epsilon^{n}}D(\epsilon,n)\Big\vert_{\epsilon=0} .
\end{equation}
Expanding the $n$-th power in \eref{eq:D_def} we can also obtain the
moments as
\begin{equation} 
  \label{newmomeqn}
  m_{n} = \frac{1}{(\rmi\mu)^{n}n!}
  \frac{\rmd^{n}}{\rmd\epsilon^{n}} \sum_{k=1}^{n}
  \left(-1\right)^{n-k} {n\choose k} 
  C(\epsilon,k)\Big\vert_{\epsilon=0} ,
\end{equation}
in terms of the correlation functions
\begin{equation} 
  \label{Cetaneqn}
  C(\epsilon,n) = \frac{1}{M}\Tr\left[
    S^{\dagger}\left(E-\frac{\epsilon\hbar\mu}{2}\right)
    S\left(E+\frac{\epsilon\hbar\mu}{2}\right)\right]^n .
\end{equation}

The correlation functions $C(\epsilon,n)$ have applications outside the
scope of this article (see \cite{kuipersetal09}) and thus their semiclassical
evaluation is a question of stand-alone interest.  However, we found
that performing the summation in \eref{newmomeqn} is a difficult
task.  Thus from now on we will work with $D(\epsilon,n)$ directly.
However, the technique we develop for evaluating $D(\epsilon,n)$ is
suitable for the simpler function $C(\epsilon,n)$ as well.  We exploit
it by deriving an equation for the generating function of
$C(\epsilon,n)$ in \ref{sec:C_all_orders}.

Finally, we note that one can use the semiclassical approximation directly
in the definition of the Wigner-Smith matrix,
equation~(\ref{eq:Q_def}).  The approach we took, however, allows us
to build on previous semiclassical work performed in open systems, in
particular on the average conductance of a chaotic ballistic device
and its moments \cite{heusleretal06,braunetal06,mulleretal07,bhn08}.
We will obtain, as in the case of the conductance and shot noise,
simple diagrammatic rules for the semiclassical contributions of
correlated trajectories.  But before approaching this task we will
quickly review the RMT results for the delay times.

\subsection{Random matrix predictions} \label{rmtpredictions}

The random matrix result for the probability distribution of the 
delay times is \cite{bfb99}
\begin{equation} \label{rmttaudisteqn}
\rho(\tau)=\frac{1}{2\pi\tau^2}\sqrt{\left(\tau_{+}-\tau\right)
\left(\tau-\tau_{-}\right)} , 
\qquad \tau_{\pm}=\frac{3\pm\sqrt{8}}{\mu} ,
\end{equation}
from which we can calculate the moments
\begin{equation}
  m_{n}
  =\int \tau^n \rho(\tau) \: \rmd\tau
  = \frac{1}{2\pi}\int_{\tau_{-}}^{\tau_{+}}\tau^{n-2}
  \sqrt{\left(\tau_{+}-\tau\right)\left(\tau-\tau_{-}\right)} \:\rmd\tau .
\end{equation}
Substituting $t+3=\mu\tau$ this can be written more obviously 
in terms of the moments of the semicircle distribution
\begin{equation} \label{firstmneqn}
m_{n}=\frac{2}{\mu^{n}}\int_{-R}^{R}(t+3)^{n-2}\frac{2\sqrt{R^2-t^2}}{\pi R^2} \: \rmd t,
\end{equation}
with $R^2=8$.  The odd moments of the semicircle distribution are 0, 
while the even moments can be written as
\begin{equation}
  \tilde{m}_{2n} = \int_{-R}^{R}t^{2n}
  \frac{2\sqrt{R^2-t^2}}{\pi R^2} \: \rmd t
  =\left(\frac{R}{2}\right)^{2n} c_{n} = 2^n c_n ,
\end{equation}
where $c_{n}$ are the Catalan numbers defined as
\begin{equation}
c_{n}=\frac{1}{n+1}\left(\begin{array}{c} 2n \\ n \end{array}\right)=\frac{(2n)!}{(n+1)!n!} .
\end{equation}
By expanding the term $(t+3)^{n-2}$ in \eref{firstmneqn} in powers of $t$, 
we can therefore express
the moments (beyond the first) by the following sums 
\begin{eqnarray}
  \label{eq:schrod_even}
  m_{2n+2} &= \frac{2}{\mu^{2n+2}} 
  \sum_{k=0}^{n} {2n \choose 2k} 3^{2(n-k)}2^{k}c_{k} , \\
  \label{eq:schrod_odd}
  m_{2n+3} &= \frac{2}{\mu^{2n+3}}
  \sum_{k=0}^{n} {2n+1 \choose 2k} 3\cdot3^{2(n-k)}2^{k} c_{k}.
\end{eqnarray}
The sequence of integers obtained from combining the Catalan numbers
as above, and hence appearing in the moments, can then be shown \cite{bv88} to be
the so-called Schr\"oder numbers (Sloane's A006318 or 2$\times$A001003) \cite{sloanes}.  More importantly for our purposes, the Schr\"oder numbers can be derived from a generating
function \cite{sloanes} and so the moments of the delay times
can also be generated from
\begin{equation} \label{rmtgenfunction}
G(s)=\sum_{n=1}\mu^{n}s^{n}m_{n}=\frac{1-s-\sqrt{1-6s+s^2}}{2} .
\end{equation}
%

\section{Calculation of the low correlation functions}
\label{firstmoments}

Since the matrix $S(E)$ is unitary we can rewrite the correlation
function $D(\epsilon,n)$ as
\begin{eqnarray} 
  \label{eq:D_alt_def}
\fl  D(\epsilon,n) &= \frac{1}{M}\Tr\left[
    S^{\dagger}\left(E-\frac{\epsilon\hbar\mu}{2}\right)
    S\left(E+\frac{\epsilon\hbar\mu}{2}\right)
    - S^{\dagger}(E)S(E)\right]^n \\
\fl  \label{eq:D_trace_exp}
  &= \frac{1}{M} \sum_{i_1,\ldots, i_n}
  \left(S^{\dagger}_{-}S_{+}-S^{\dagger}S\right)_{i_1,i_2}
  \left(S^{\dagger}_{-}S_{+}-S^{\dagger}S\right)_{i_2,i_3}\cdots ,
\end{eqnarray}
where the summation in the second line is over choices of $n$ incoming channels and 
the $\pm$ subscripts represent the different energy arguments of the scattering matrices.
To evaluate $D(\epsilon,n)$ semiclassically, we will use the
semiclassical approximation for the scattering matrix elements, which
is given in terms of open trajectories \cite{miller75,richter00,rs02}
\begin{equation} 
  \label{scatmateqn}
  S_{ba}(E) \approx \frac{1}{\sqrt{T_{\mathrm{H}}}}\sum_{\zeta (a \to b)}
  A_{\zeta}\rme^{\frac{\rmi}{\hbar}S_{\zeta}} .
\end{equation}
The $T_{\mathrm{H}}$ appearing in the prefactor is the Heisenberg
time, and it is simply related to the classical escape rate by $\mu =
M/T_{\mathrm{H}}$.  In the sum which is over all classical
trajectories $\zeta$ that start in channel $a$ and end in channel $b$
(where the channels fix the absolute value of the angles at which the
trajectories enter and leave the cavity) $S_{\zeta}$ is the action of
the trajectory $\zeta$ and $A_\zeta$ its stability amplitude
\cite{richter00} including the phase due to the number of conjugate
points along the trajectory.  As our starting point, we will
substitute approximation \eref{scatmateqn} into \eref{eq:D_alt_def},
expand the action up to first order in energy $S_{\zeta}(E+\delta
E)\approx S_{\zeta}(E) + T_{\zeta}(E)\delta E$, where $T_{\zeta}$ is
the time the trajectory $\zeta$ spends inside the system, and ignore
any change in the slowly varying prefactor $A_\zeta$.

Below we perform this calculation for $n=1$, $2$ and $3$ before
formulating the general counting rules and performing the evaluation
of $D(\epsilon,n)$ for general $n$.

\subsection{Calculating $D(\epsilon,1)$}

After semiclassical approximation \eref{scatmateqn}, the
correlation function $D(\epsilon,1)$ becomes
\begin{eqnarray} 
  \label{cepssemi}
\fl  D(\epsilon,1) &= \frac{1}{M} \sum_{i_1}
  \left(S^{\dagger}_{-}S_{+}-S^{\dagger}S\right)_{i_1,i_1}   \nonumber\\ 
\fl  & \approx \frac{1}{M T_{\mathrm{H}}}\sum_{i_1,o_1}
  \sum_{\zeta, \zeta'(i_1\to o_1)}A_{\zeta}A^{*}_{\zeta'}
  \rme^{\frac{\rmi}{\hbar}(S_{\zeta}-S_{\zeta'})}
  \left(\rme^{\frac{\rmi \epsilon \mu}{2} (T_{\zeta}+T_{\zeta'})} - 1\right) ,
\end{eqnarray}
which is a sum over trajectory pairs $\zeta,\zeta'$ both of which
start and end in the same channels ($i_1$ and $o_1$ respectively),
followed by a sum over all the possible channels.  We note that the
difference of actions $S_{\zeta}-S_{\zeta'}$ is divided by $\hbar \ll
1$ and therefore the resulting phase oscillates wildly unless the
action difference is of order $\hbar$.  The semiclassical expansion is
based on identifying couples (or, more generally, families) of orbits
that have small action differences. 

To leading order in inverse channel number $1/M$ the first
correlation function can be calculated as in \cite{ks08} using the
diagonal approximation \cite{berry85}.  This approximation restricts
the sum to trajectories $\zeta$ and $\zeta'$ that are identical
\begin{equation} 
  \label{diag1}
  D^{\mathrm{diag}}(\epsilon,1) = \frac{1}{M T_{\mathrm{H}}} 
  \sum_{i_1,o_1} \sum_{\zeta (i_1\to o_1)} \vert A_{\zeta}\vert^{2}
  \left(\rme^{\rmi \epsilon \mu T_{\zeta}} - 1\right) .
\end{equation}
The sum in \eref{diag1} can be performed by using a sum rule for open
trajectories \cite{rs02} which turns it into an integral over the
trajectory time $T$
\begin{equation} 
  \label{opensumruleeqn}
  \sum_{\zeta(i_1\to o_1)}\vert A_{\zeta}\vert^{2} \ldots
  \approx \int_{0}^{\infty}\rmd T \: \rme^{-\mu T} \ldots .
\end{equation}
The exponential term in \eref{opensumruleeqn} represents the average
probability that a trajectory remains in the system for the time $T$,
while the sum over channels, where we can pick both $i_1$ and $o_1$ from
the $M$ possible channels, simply gives a factor of $M^{2}$.  The
diagonal approximation thus gives
\begin{eqnarray}
  D^{\mathrm{diag}}(\epsilon,1) &\approx \frac{M}{T_{\mathrm{H}}}
  \int_{0}^{\infty}\rmd T \: 
  \left(\rme^{-\mu (1 - \rmi\epsilon)T} - \rme^{-\mu T}\right) \\
  &= \frac{1}{1 - \rmi \epsilon} - 1 
  = \frac{\rmi \epsilon}{1 - \rmi \epsilon},
\end{eqnarray}
where we have used the fact that $\mu = M/T_{\mathrm{H}}$ to simplify.
Substituting into \eref{eq:mom_thru_D} we obtain
\begin{equation}
  \tau_{\mathrm{W}}  = \frac{1}{\mu} ,
\end{equation}
which just states that the average delay time is the inverse of the
classical escape rate as we might expect.

\subsection{Calculating $D(\epsilon,2)$}

To obtain the next moment we move to the correlation function
$D(\epsilon,2)$.  For this we will need to treat not only the diagonal
pairs, but also correlated trajectories that have encounters.  The
calculation follows the calculation of the shot noise power
\cite{braunetal06} which was first performed to leading order in inverse 
channel number for quantum graphs \cite{spg03}.  The semiclassical treatment 
of the shot noise builds on work on the conductance
\cite{rs02,heusleretal06}, which itself is built on the work on spectral
statistics \cite{sr01,mulleretal04,mulleretal05}.

First we slightly modify (\ref{eq:D_trace_exp}) to explicitly include
an indication of the unitarity of $S$
\begin{equation}
 \label{eq:D2_trace_exp}
\fl   D(\epsilon,2) = \frac{1}{M} \sum_{i_1, i_2} 
    \left(S^{\dagger}_{-}S_{+} 
      - \delta_{i_1,i_2}S^{\dagger}S \right)_{i_1,i_2}
    \left(S^{\dagger}_{-}S_{+} 
      - \delta_{i_1,i_2}S^{\dagger}S\right)_{i_2,i_1} .
\end{equation}
While the semiclassical approximation (\ref{scatmateqn}) preserves the
unitarity of $S$, and thus the Kronecker deltas are not necessary,
their inclusion greatly facilitates the derivation.

We now write down the semiclassical expression for the
correlation function $D(\epsilon,2)$ in terms of open trajectories
\begin{eqnarray} 
  \label{eq:D2traj}
  \fl D(\epsilon,2) \approx \frac{1}{M T_{\mathrm{H}}^{2}}
  \sum_{\substack{{i_1,i_2} \cr {o_1,o_2}}} 
  \sum_{\substack{\zeta(i_1\to o_1) \cr 
      \zeta' (i_2\to o_1)}}
  \sum_{\substack{\xi(i_2\to o_2) \cr \xi' (i_1\to o_2)}}
  A_{\zeta}A_{\zeta'}^{*}A_{\xi}A_{\xi'}^{*}
  \rme^{\frac{\rmi}{\hbar}(S_{\zeta}-S_{\zeta'}+S_{\xi}-S_{\xi'})}  \nonumber \\
  \times 
  \left(\rme^{\frac{\rmi \epsilon \mu}{2} (T_{\zeta}+T_{\zeta'})}
    - \delta_{i_1,i_2}\right)
  \left(\rme^{\frac{\rmi \epsilon \mu}{2} (T_{\xi}+T_{\xi'})}
    - \delta_{i_1,i_2}\right) . 
\end{eqnarray}
For diagonal terms we can either pair $\zeta=\zeta'$ and $\xi=\xi'$, or
$\zeta=\xi'$ and $\xi=\zeta'$.  For the first case, the start channels
$i_1$ and $i_2$ must coincide thus triggering the Kronecker delta.
For the second case we have $o_1=o_2$.  Either case leads to a
(leading order) channel factor of $M^3$, because the number of
channels we can choose in the outer summation is 3.  Putting these two
possibilities into (\ref{eq:D2traj}) we can write the diagonal
contribution as
\begin{eqnarray}
  D^{\mathrm{diag}}(\epsilon,2) = &\frac{M^{2}}{T_{\mathrm{H}}^{2}}
  \sum_{\zeta, \xi}|A_{\zeta}|^{2} |A_{\xi}|^{2}
  \left(\rme^{{\rmi \epsilon \mu} T_{\zeta}} - 1\right)
  \left(\rme^{{\rmi \epsilon \mu} T_{\xi}} - 1\right) \nonumber \\
  & + \frac{M^{2}}{T_{\mathrm{H}}^{2}} 
  \sum_{\zeta, \xi} |A_{\zeta}|^{2} |A_{\xi}|^{2}
  \rme^{{\rmi \epsilon \mu} T_{\zeta}}
  \rme^{{\rmi \epsilon \mu} T_{\xi}} .
\end{eqnarray}

Note that in the second case it can also happen that $i_1=i_2$, but
this case gives a lower order channel factor of $M^2$ and is therefore
neglected.  Using the open sum rule from \eref{opensumruleeqn} we get
\begin{equation} 
  \label{eq:D2_diag_ans}
  D^{\mathrm{diag}}(\epsilon,2) 
  = \frac{(\rmi \epsilon)^2}{\left(1-\rmi\epsilon\right)^2}
  + \frac{1}{\left(1-\rmi\epsilon\right)^2} .
\end{equation}

\begin{figure}
  \centering
  \includegraphics[width=8cm]{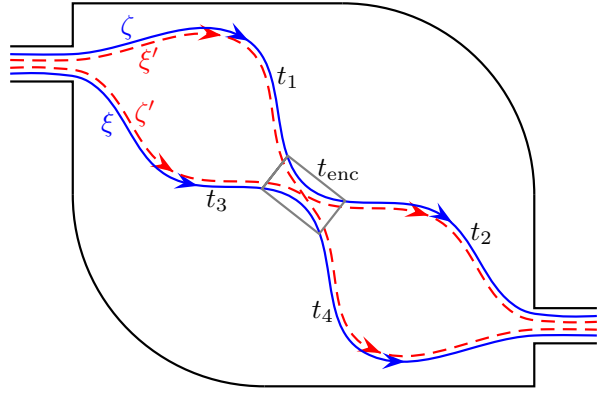}
  \caption{An example of two trajectories with a single encounter and
    two partner trajectories.}
  \label{trajectoryquad}
\end{figure}

However, as we know from the calculation of the shot noise
\cite{spg03,braunetal06}, the diagonal terms are
not the only ones that contribute to leading order in inverse channel
number.  If the trajectories $\zeta$ and $\xi$ come very close to each
other in an encounter, as in Figure~\ref{trajectoryquad}, then the
partner trajectories can cross over inside the encounter leading to a
quadruplet of trajectories with a small action difference.  Such a
quadruplet can then give a contribution in the semiclassical limit.
While such an encounter makes the contribution higher order in inverse
channel number, the sum over channels now contributes the factor of
$M^{4}$.  As a result the quadruplet contributes at the same order as
the diagonal terms.

To calculate the contribution, we simply put the additional energy
arguments into the calculation of the shot noise power.  The
contribution can be separated into a product over the links and the
encounters \cite{braunetal06} and written as
\begin{eqnarray}
  D^{(2^1)}(\epsilon,2)
  = &\frac{M^{3}}{T_{\mathrm{H}}^{2}}
  \int_{0}^{\infty}\rmd t_{1}\:\rme^{-\mu(1-\rmi\epsilon)t_1}
  \int_{0}^{\infty}\rmd t_{2}\:\rme^{-\mu(1-\rmi\epsilon)t_2}
  \nonumber \\
  & \times \int_{0}^{\infty}\rmd t_3\:\rme^{-\mu(1-\rmi\epsilon)t_3}
  \int_{0}^{\infty}\rmd t_{4}\:\rme^{-\mu(1-\rmi\epsilon)t_4} 
  \nonumber \\
  \label{eq:D2_x}
  &\times \int \rmd s \rmd u \: 
  \frac{\rme^{-\mu(1-2\rmi \epsilon)t_{\mathrm{enc}}(s,u)} 
    \rme^{\frac{\rmi}{\hbar}su}}{\Omega t_{\mathrm{enc}}(s,u)} ,
\end{eqnarray}
where $(2^1)$ refers to the structure of the diagram: one encounter
with two (unprimed) trajectories meeting (a ``2-encounter'').  In the
final integral, $s$ and $u$ are the separations along the stable and
unstable manifolds of the two original stretches inside the encounter
and $\Omega$ is the volume of the available phase space, while
$t_{\mathrm{enc}}(s,u)$ is the duration of the encounter.  An
important point is that although the encounter involves two trajectory
stretches, as they are close to each other they will either remain
inside the system or escape together and their average survival
probability is given by the time of just a single stretch.  The
presence of the encounter actually slightly enhances the survival
probability of the whole trajectory quadruplet, and this tiny
classical effect has important semiclassical implications.  Elsewhere
in \eref{eq:D2_x}, the $t_i$ are the durations of the link stretches
as depicted in Figure~\ref{trajectoryquad}.  Performing the integrals
following \cite{braunetal06}, we obtain the result
\begin{equation} 
  \label{eq:D2_x_ans}
  D^{(2^1)}(\epsilon,2) = \frac{-(1-2\rmi\epsilon)}{(1-\rmi\epsilon)^4},
\end{equation}
The structure of the answer is very simple: each $\enc$-encounter
contributes a factor of $-\left(1-\rmi \enc \epsilon\right)$, while
each link stretch gives the factor $\left(1-\rmi\epsilon\right)^{-1}$.  These
diagrammatic rules, which first arose for the conductance
\cite{heusleretal06}, massively simplify calculating the semiclassical
contributions, and are the reason we consider correlation functions
rather than the time delay directly.

We can combine the two leading order results from \eref{eq:D2_diag_ans} and
\eref{eq:D2_x_ans} to obtain the second moment
\begin{equation} 
  \label{eq:second_mom_ans}
  m_2 = \frac{2}{\mu^2} .
\end{equation}

In conclusion of this subsection we mention that the configurations
described as diagonal contributions above can be obtained from the
diagram in Figure~\ref{trajectoryquad} by setting $t_1=t_3=0$ for the
first case and $t_2=t_4=0$ for the second.  We will refer to this
reduction as moving or sliding an encounter into the lead.  As we have seen
above, moving an encounter into the input lead can lead to a
contribution that is different from the encounter in the output lead.

\subsection{Calculating $D(\epsilon,3)$}

\begin{figure}
  \centering
  \includegraphics[width=12cm]{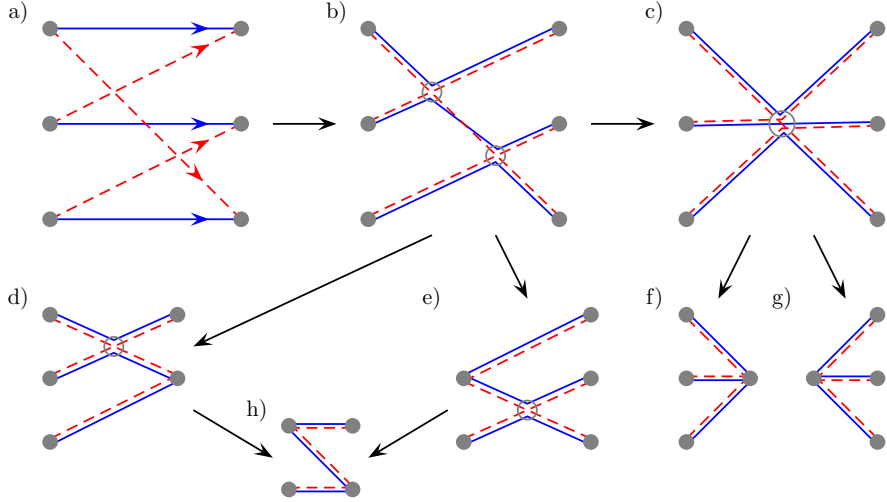}
  \caption{The original trajectory structure for $D(\epsilon,3)$ in a)
    can be collapsed down to (three copies) of the structure in b),
    which has two 2-encounters, to ensure a small action difference.
    Sliding the encounters together creates the single 3-encounter in
    c), while moving encounters into the leads generates all the
    further possibilities.  Trajectories $\zeta_j$ are indicated by
    solid lines and trajectories $\zeta_j'$ by dashed lines.}
  \label{ceps3fig}
\end{figure}

Before we proceed to calculate $D(\epsilon,3)$, we will briefly look
at how we can form the diagrams that contribute.  As we have seen, the
$n$-th correlator is expressed as a sum over $2n$ trajectories
\begin{eqnarray} 
  \label{Depsnsemieqn}
  \fl D(\epsilon,n)  \approx  \frac{1}{M {T_{\mathrm{H}}}^{n}}
  \sum_{\{i_j,o_j\}} \sum_{\substack{\left\{\zeta_j(i_j\to o_j)\right\} 
      \cr \{\zeta_{j}'(i_{j+1}\to o_j)\}}} 
  \prod_{j=1}^{n}
  A_{\zeta_{j}}A_{\zeta_{j}'}^{*} 
  \rme^{\frac{\rmi}{\hbar}(S_{\zeta_{j}} - S_{\zeta_{j}'})}
  \left(\rme^{\frac{\rmi\epsilon\mu}{2}(T_{\zeta_{j}}+T_{\zeta_{j}'})} 
    - \delta_{i_j, i_{j+1}} \right). \nonumber \\
&&
\end{eqnarray}
Taking the trace of the product of matrices means that we identify
$i_{n+1}=i_{1}$.  Therefore the trajectories complete a cycle, if we
consider moving forward along the unprimed trajectories and back along
the primed ones.  The resulting structure for $n = 3$ is shown in
Figure~\ref{ceps3fig}a and, as we have also seen, in
\eref{Depsnsemieqn} we add the actions of the unprimed trajectories
and subtract the actions of the primed ones, so the resulting phase
oscillates wildly unless the total action difference is of the order
of $\hbar$.  

To obtain such a small action difference we can collapse all the
trajectories onto each other, as in Figure~\ref{ceps3fig}b, creating
encounters of the type we saw in Figure~\ref{trajectoryquad}.  It
turns out that to obtain all contributions from this type of direct
collapse we need to cyclically permute the labels of the trajectories,
resulting in three copies of the diagram in
Figure~\ref{ceps3fig}b. Alongside this direct collapse we can imagine
sliding the encounters together to create a single diagram with a
single 3-encounter, as shown in Figure~\ref{ceps3fig}c.  Further
possibilities then arise from sliding the encounters into the leads,
giving the remaining diagrams depicted in Figure~\ref{ceps3fig}.

Before we write down the total contribution, we consider two
instructive examples.  First we evaluate the contribution from the 
diagram in Figure~\ref{ceps3fig}e.  We assume the trajectories indicated by solid lines are
numbered top to bottom as $\zeta_1$, $\zeta_2$ and $\zeta_3$.  Then we
have $i_1=i_2$ since the first encounter is in the incoming lead.
This activates one of the Kronecker deltas in the expression for
$D(\epsilon,3)$ (see \eref{Depsnsemieqn} or the similar expression in 
\eref{eq:D2_trace_exp}).  The terms due to trajectories $\zeta_2$ and $\zeta_3$,
which have a non-degenerate encounter, are the same as in
\eref{eq:D2_x}, giving in total
\begin{eqnarray}
  \label{eq:D3_d3}
  \fl D^{(2^2)}_{i_1=i_2}(\epsilon,3) = &\frac{M^4}{T_{\mathrm{H}}^3}
  \int_0^\infty \rmd T_{\zeta_1} \: \rme^{-\mu T_{\zeta_1}}
  \left(\rme^{\rmi\epsilon\mu T_{\zeta_1}} - 1\right) \prod_{j=2,3} \prod_{p=1,2}
  \int_{0}^{\infty}\rmd T_{\zeta_j}^p\:\rme^{-\mu(1-\rmi\epsilon)T_{\zeta_j}^p} \nonumber \\ 
\fl &\times 
  \int \rmd s \rmd u \: 
  \frac{\rme^{-\mu(1-2\rmi \epsilon)t_{\mathrm{enc}}(s,u)} 
    \rme^{\frac{\rmi}{\hbar}su}}{\Omega t_{\mathrm{enc}}(s,u)} ,
\end{eqnarray}
where $T_{\zeta}^p$ refers to the duration of $p$-th part of
trajectory $\zeta$ and the power of $M$ came from the 5 choices of the 
remaining channels.  Evaluating the integrals we get
\begin{equation}
  \label{eq:D3_d3_ans}
  D^{(2^2)}_{i_1=i_2}(\epsilon,3) 
  = \frac{-\rmi\epsilon(1-2\rmi\epsilon)}{(1-\rmi\epsilon)^5} .
\end{equation}

If on the other hand, we consider the contribution of the diagram in 
Figure~\ref{ceps3fig}g, we notice that
$i_1=i_2=i_3$, thus activating three Kronecker deltas.  The
contribution of this diagram is thus
\begin{equation}
  \label{eq:D3_d5}
  \fl D^{(3^1)}_{i_1=i_2=i_3}(\epsilon,3) = \frac{M^3}{T_{\mathrm{H}}^3}
  \prod_{j=1}^3 \int_0^\infty \rmd T_{\zeta_j} \: \rme^{-\mu T_{\zeta_j}}
  \left(\rme^{\rmi\epsilon\mu T_{\zeta_j}} - 1\right)
  = \frac{(\rmi\epsilon)^3}{(1-\rmi\epsilon)^3} .
\end{equation}

Comparing the results in \eref{eq:D2_diag_ans}, \eref{eq:D3_d3_ans}
and \eref{eq:D3_d5} we can surmise that the power of $\rmi\epsilon$
that results from sliding an encounter into the incoming lead is equal to
the number of {\em direct\/} stretches from the encounter to the
outgoing lead, i.e.~the stretches that do not participate in any other
encounters.  This observation can be mathematically verified
by using the relationship between diagrams and factorisations of the
cyclic permutation \cite{bhn08b}.

To summarise, we can write down the contribution of each diagram by
simply looking at its links and encounters and assigning
\begin{itemize}
\item a factor of $(1-\rmi\epsilon)^{-1}$ to each stretch,
\item a factor of $-(1-\rmi\enc\epsilon)$ to each non-degenerate
  $\enc$-encounter,
\item a factor of $(\rmi\epsilon)^s$ to each encounter happening in
  the incoming lead and having $s$ {\em direct\/} stretches to the
  outgoing lead.
\item a factor of 1 to each encounter happening in the outgoing lead.
\end{itemize}
Altogether, the leading contribution to $D(\epsilon,3)$ is thus
\begin{eqnarray}
  \label{eq:D3_tot_ans}
  \fl D(\epsilon,3) = & 3 \left(
    \frac{(1-2\rmi\epsilon)^2}{(1-\rmi\epsilon)^7}
    + \frac{-(1-2\rmi\epsilon)}{(1-\rmi\epsilon)^5}
    + \frac{-\rmi\epsilon(1-2\rmi\epsilon)}{(1-\rmi\epsilon)^5}
    + \frac{\rmi\epsilon}{(1-\rmi\epsilon)^3} \right)
  \nonumber \\ 
\fl  & {} + \frac{-(1-3\rmi\epsilon)}{(1-\rmi\epsilon)^6}
  + \frac{1}{(1-\rmi\epsilon)^3}
  + \frac{(\rmi\epsilon)^3}{(1-\rmi\epsilon)^3} ,
\end{eqnarray}
where the factor of three counts the three different ways to label the
diagram of Figure~\ref{ceps3fig}b and its descendants.

The third moment can thus be calculated to be
\begin{equation}
  m_{3} = \frac{6}{\mu^3} ,
\end{equation}
and we note that if we put in three different energy arguments in line 
with \eref{eq:Cvecepsn}, we can use the same diagrams to get the 
next three moments using \eref{eq:odd_mom} and \eref{eq:even_mom} as
\begin{equation}
  \label{eq:higher_low_moments}
  m_4 = \frac{22}{\mu^4} ,
  \qquad m_5 = \frac{90}{\mu^5} ,
  \qquad m_6 = \frac{394}{\mu^6} .
\end{equation}
%

\section{All moments} \label{allmoments}

\begin{figure}
  \centering
  \includegraphics{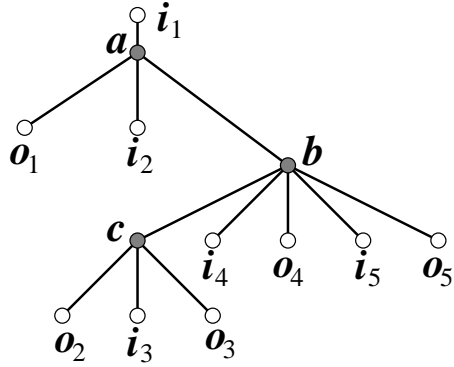}
  \caption{An example of a $(2^23^1)$-tree.}
  \label{fig:tree_example}
\end{figure}

Now that we know the rules which govern the contributions of
individual diagrams, we can look at generating all the diagrams and
their contributions recursively.  To leading order in inverse channel
number, the contributing diagrams of the type $2^{v_2}3^{v_3}\cdots$
are in a bijective correspondence with rooted plane trees
\cite{bhn08} that have $v_\enc$ vertices of degree $2\enc$ and all
other vertices of degree one (called ``leaves''), see
Figure~\ref{fig:tree_example}.  We denote by $V=v_2+v_3+\ldots$ the
total number of vertices of degree higher than 1.  These vertices
correspond to encounters in the diagram.  The total number of leaves
can easily be seen to be $2(L-V+1)$, where $L=2v_2 + 3v_3 + \ldots$.
Starting from the root the leaves are labelled $i_1$, $o_1$, $i_2$,
$o_2$, \ldots, $i_n$, $o_n$, where $n=L-V+1$ is the order of the
correlation function $D(\epsilon,n)$.  The $i$-labelled leaves
correspond to incoming trajectories starting in the lead and the
$o$-labelled leaves correspond to the outgoing trajectories exiting
into the lead.

As we have seen, some encounters can touch the lead, but this can only
happen if an $\enc$-encounter has $\enc$ vertices with labels $i$
connected to it ($i$-touch) or $\enc$ vertices with labels $o$
connected to it ($o$-touch).  For example, in
Figure~\ref{fig:tree_example}, the vertex $a$ can $i$-touch, the
vertex $c$ can $o$-touch and the vertex $b$ can do neither.

\begin{figure}
  \centering
  \includegraphics[width=12cm]{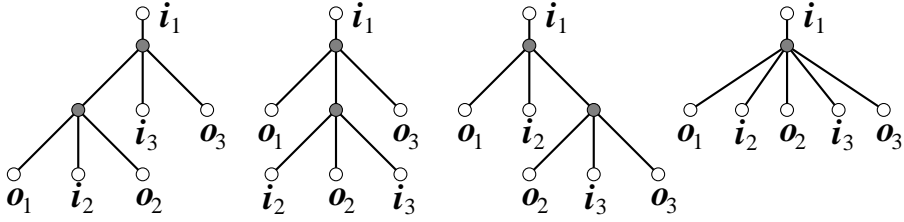}
  \caption{The trees contributing to the third order correlation
    function $D(\epsilon,3)$.  The first three trees correspond to
    relabellings of Figure~\ref{ceps3fig}b and the last one to
    Figure~\ref{ceps3fig}c.}
  \label{fig:oreder3trees}
\end{figure}

When an encounter {\em can\/} touch the lead, the total answer we seek
is the sum of the contributions from when it {\em does\/} and when it
{\em does not\/} do it.  Equivalently, we can take the multiplicative
factor of an encounter to be the sum of all factors it can produce.
To illustrate this point, we revisit the calculation of
$D(\epsilon,3)$ and rewrite equation \eref{eq:D3_tot_ans} in the form
\begin{eqnarray}
  \label{eq:D3_tot_ans_redux}
  D(\epsilon,3) =& \frac{3}{(1-\rmi\epsilon)^3}
  \left(\frac{-(1-2\rmi\epsilon)}{(1-\rmi\epsilon)^2} + 1\right)
  \left(\frac{-(1-2\rmi\epsilon)}{(1-\rmi\epsilon)^2} 
    + \rmi\epsilon \right)
  \nonumber \\
  &+ \frac{1}{(1-\rmi\epsilon)^3}
  \left(\frac{-(1-3\rmi\epsilon)}{(1-\rmi\epsilon)^3} 
    + 1 + (\rmi\epsilon)^3\right) .
\end{eqnarray}
This is the sum of contributions of three $(2^2)$-trees and one
$(3^1)$-tree, see Figure~\ref{fig:oreder3trees}.  The structure of a
contribution is as follows: the prefactor is $(1-\rmi\epsilon)^{-n}$,
where $n$ is half the number of leaves (the order of the correlation
function).  Then follow the factors corresponding to the vertices of
the diagram.  Each vertex of degree $2\enc$ (corresponding to an
$\enc$-encounter) gives a multiplicative factor of
\begin{equation}  
  \label{eq:mult_normal}
  \frac{-(1 - \rmi \enc \epsilon)}{(1-\rmi \epsilon)^{\enc}} ,
\end{equation}
modified by an additional $+1$ if the vertex can $o$-touch and by
$+(\rmi\epsilon)^s$ if the vertex can $i$-touch, where $s$ is the
number of $o$-leaves attached to the vertex.  In the case when there
is only one vertex in the diagram, as in the $(3^1)$-tree in
Figure~\ref{fig:oreder3trees}, it can both $i$-touch and $o$-touch but
not at the same time.  To provide a further example, the overall
contribution of the tree in Figure~\ref{fig:tree_example} is
\begin{equation}
  \frac{1}{(1-\rmi \epsilon)^5}
  \left( \frac{-(1 - 2\rmi \epsilon)}{(1-\rmi \epsilon)^2} 
    + \rmi \epsilon \right)
  \left(\frac{-(1 - 3 \rmi \epsilon)}{(1-\rmi \epsilon)^3}\right)
  \left( \frac{-(1 - 2\rmi \epsilon)}{(1-\rmi \epsilon)^2} + 1 \right) .
\end{equation}

To count all possible trees while keeping track of the structure of
their vertices we introduce the generating function $F(\vec{x},
\vec{z_o}, \sigma, \vec{z_i}, \tau)$.  The roles of the variables are
as follows:
\begin{itemize}
\item the power of $x_\enc$ enumerates the number of
  non-degenerate $\enc$-encounters
\item the power of $z_{o, \enc}$ enumerates the number of $\enc$-encounters
  that $o$-touch the lead
\item the power of $1+\sigma$ is the total number of $i$-labelled leaves
  adjacent to the encounters that $o$-touch the lead
\item the power of $z_{i, \enc}$ enumerates the number of $\enc$-encounters
  that $i$-touch the lead
\item  the power of $1+\tau$ is the total number of $o$-labelled leaves
  adjacent to the encounters that $i$-touch the lead.
\end{itemize}
For example, the tree in Figure~\ref{fig:tree_example} gives rise to
four contributions to the generating function, corresponding to the four
possibilities of the vertices $a$ and $c$ touching the lead or not
\begin{eqnarray}
  \label{eq:gen_fun_example}
  \fl  x_2 x_3 x_2 + z_{i,2}(1+\tau) x_3 x_2 + x_2 x_3 z_{o,2}(1+\sigma) +
  z_{i,2}(1+\tau) x_3 z_{o,2}(1+\sigma) 
  \nonumber \\
  = \Big(x_2+z_{i,2}(1+\tau)\Big) x_3 \Big(x_2 + z_{o,2}(1+\sigma)\Big) .
\end{eqnarray}
Our aim then is to set
\begin{equation}
  \label{eq:var_vals}
  \fl  x_\enc = \frac{-(1 - \rmi \enc \epsilon)}{(1-\rmi
    \epsilon)^{\enc}}, 
  \qquad z_{o,\enc} = z_{i,\enc} = 1, \qquad \sigma = 0, 
  \qquad \tau=-1+i\epsilon ,
\end{equation}
in line with the semiclassical contributions described above, and
finally include a change of variables to provide the correct
prefactor of $(1-\rmi\epsilon)^{-n}$.
\begin{figure}
  \centering
  \includegraphics[scale=0.7]{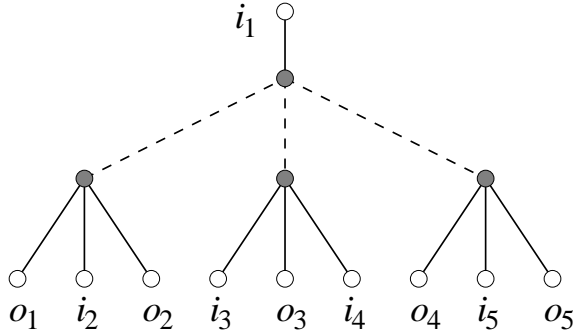}
  \caption{A separation of a tree into the top vertex and subtrees.
    Note that in the middle subtrees the roles of variables $(\vec{z_o},
    \sigma)$ and $(\vec{z_i}, \tau)$ are reversed.}
  \label{fig:separation}
\end{figure}

While the generating function $F$ is our aim, we will first deal with
an auxiliary function $f = f(\vec{x}, \vec{z_o}, \sigma, \vec{z_i},
\tau)$ which is defined exactly as $F$ except for not allowing the top
vertex to $i$-touch the lead and not counting the root as a leaf when
the top vertex $o$-touches the lead.  These restrictions on the
function $f$ makes it possible to find a recursive equation for it.
The value corresponding to an empty tree will be set to 1,
i.e.~$f(0)=1$.  The correlation function calculated with the above
restrictions will be denoted by $\tilde{D}(\epsilon,n)$.

To write a recursion for $f$ we separate a tree into the top vertex of
degree $2\enc$ and $2\enc - 1$ subtrees, see
Figure~\ref{fig:separation}.  If the top vertex is non-degenerate, its
contribution is $x_\enc f^\enc \fh^{\enc-1}$, where the function
$\hat{f}$ correspond to the even-numbered subtrees in which the
positions of $i$'s and $o$'s are reversed.  Thus the roles of all the
variables corresponding to leaves of one type are switched, i.e.~$\fh
= f(\vec{x}, \vec{z_i}, \tau, \vec{z_o}, \sigma)$.

To account for the possibility of the top vertex touching the lead, we
recall that from the definition of $f$, it is only allowed to
$o$-touch.  In this case all odd-numbered subtrees must be empty and
the contribution of each even-numbered subtree is $\fh + \tau$.
Note that if the subtree is empty, the contribution evaluates to the
correct value of $\fh(0)+\tau = 1+\tau$.  Putting this together,
we have
\begin{equation}
  \label{eq:recur_f}
  f = 1 + \sum_{\enc=2}^\infty \left[ x_\enc f^{\enc}
    {\fh}^{\enc-1} + z_{o,\enc} (\fh + \tau)^{\enc-1} \right] ,
\end{equation}
and correspondingly
\begin{equation}
  \label{eq:recur_fhat}
  \fh = 1 + \sum_{\enc=2}^\infty \left[ x_\enc {\fh}^{\enc}
    {f}^{\enc-1} + z_{i,\enc} (f + \sigma)^{\enc-1} \right] .
\end{equation}
To obtain the function $F$ we need to take into account the special
role of the top vertex.  It can both $i$-touch and $o$-touch although not
at the same time.  Additionally, it is always adjacent to an
$i$-labelled leaf and thus always contributes to the power of
$1+\tau$.  The final generating function $F$ satisfies
\begin{eqnarray}
  \nonumber
\fl F &= 1 + \sigma + \tau + \sum_{\enc=2}^\infty \left[
    x_\enc f^{\enc} {\hat{f}}^{\enc-1}
    + z_{o,\enc} (\hat{f} + \tau)^{\enc-1}(1+\tau)
    + z_{i,\enc} (f+\sigma)^{\enc}  \right] \\
  \nonumber
\fl &= f + \sigma + \tau 
  + \sum_{\enc=2}^\infty \left[\tau z_{o,\enc} (\hat{f} + \tau)^{\enc-1}
    + z_{i,\enc} (f+\sigma)^{\enc} \right] \\
  \label{eq:recur_F}
\fl &= \sum_{l=1}^\infty \left[\tau z_{o,\enc} (\hat{f} + \tau)^{\enc-1}
    + \sigma z_{i,\enc} (f+\sigma)^{\enc-1} 
    + f z_{i,\enc} (f+\sigma)^{\enc-1} 
  \right],
\end{eqnarray}
where, to get to the last line we defined $z_{o,1} = z_{i,1} = 1$.  We
note that the value $F(0)$ cannot be defined from recursive
considerations and needs to be chosen to provide the correct answer
for $D(\epsilon,1)$, which turns out to be $\rmi\epsilon =
1+\sigma+\tau$.  Another important observation is that our choice of
the leaf $i_1$ as the root (see Figure~\ref{fig:tree_example} for example)
is arbitrary.  In particular an $o$-leaf could be chosen and the answer
for $D(\epsilon,n)$ should not depend on the choice.  Thus the
function $F$ should be symmetric with respect to swapping the
variables $z_{o}$ with $z_i$, $\tau$ with $\sigma$ and $f$ with
$\hat{f}$.  This is not apparent from (\ref{eq:recur_F}) but will be
checked (and used!) at a later stage.

Now we can make the substitutions
\begin{eqnarray}
  \label{eq:subss}
  x_\enc &= -\frac{1 - \rmi \enc\epsilon}{(1-\rmi\epsilon)^\enc}\, 
  \tilde{r}^{\enc-1} ,
  \\
  z_{o,\enc} &= z_{i,\enc} = \tilde{r}^{\enc-1} , \\
  \label{eq:subs_r}
   \tilde{r} &= \frac{r}{(1-\rmi\epsilon)}, \\
  \label{eq:substausigma} 
  \sigma &= 0, \qquad \tau=-1+\rmi\epsilon.
\end{eqnarray}
The substitutions give the contribution of each tree diagram as in
\eref{eq:var_vals}, but we have included the powers of $\tilde{r}$ to
keep track of which order correlation function they contribute to.
Indeed, the power of $\tilde{r}$ corresponding to a
$2^{v_2}3^{v_3}\cdots$ tree would be $v_2(2-1) + v_3(3-1) + \ldots =
L-V = n-1$.  Substitution \eref{eq:subs_r} therefore gives a prefactor
of $(1-\rmi\epsilon)^{1-n}$ and so to get the additional factor of
$(1-\rmi\epsilon)^{-1}$ we need for the correct prefactor, we
introduce $g = (1-\rmi\epsilon)^{-1} f$.  The function $g$ is a 
generating function of the ``restricted'' coefficients
$\tilde{D}(\epsilon,n)$, i.e.
\begin{equation}
  \label{eq:g_meaning}
  g = \sum_{n=1}^\infty r^{n-1} \tilde{D}(\epsilon,n).
\end{equation}
Performing all the changes of variable, apart from \eref{eq:substausigma} for
now, from \eref{eq:recur_f} we arrive at
\begin{equation}
  \label{eq:recur_g}
\fl  g(1-\rmi\epsilon) = 1 - \sum_{\enc=2}^\infty r^{\enc-1}g^{\enc}\gh^{\enc-1}(1-\rmi \enc\epsilon) + \sum_{\enc=2}^\infty r^{\enc-1}\left(\gh + \frac{\tau}{(1-\rmi\epsilon)}\right)^{\enc-1} ,
\end{equation}
and a similar equation for $\gh$.  The sums can be performed easily,
especially when we notice that the first two terms correspond to the
$\enc=1$ terms of the two sums, leading to
\begin{equation}
\frac{g}{1-rg\gh} = \frac{\rmi\epsilon g}{(1-rg\gh)^2} + \frac{1}{1-r\gh -\frac{r\tau}{(1-\rmi\epsilon)}} .
\end{equation}
Taking the numerator of the above equation, we arrive at
\begin{equation}
 \label{eq:firstgimplicit}
  (1-\rmi\epsilon-rg\gh)\left[\frac{1+rg\tau}{1-\rmi\epsilon}-g\right]+(\rmi\epsilon)^2rg\gh = 0 ,
\end{equation}
and similarly for $\gh$
\begin{equation}
 \label{eq:firstghimplicit}
  (1-\rmi\epsilon-rg\gh)\left[\frac{1+r\gh\sigma}{1-\rmi\epsilon}-\gh\right]+(\rmi\epsilon)^2rg\gh = 0 .
\end{equation}
We have written the equations in a form that highlights the
symmetric terms involving $g\gh$.  Taking the difference between
\eref{eq:firstgimplicit} and \eref{eq:firstghimplicit} we obtain
\begin{equation}
  \label{eq:rel_g_ghat}
  g(1-\rmi\epsilon-r\tau) = \gh(1-\rmi\epsilon-r\sigma) .
\end{equation}
This can now be substituted back into \eref{eq:firstgimplicit} or
\eref{eq:firstghimplicit} to give an implicit equation for $g$ or
$\gh$, although we only state the result after the simplifying 
substitution from \eref{eq:substausigma}
\begin{equation} \label{gnewgenfunctioneqn}
\left[g-\frac{1}{\left(1-\rmi\epsilon\right)}\right]\left(1+r\right)=\frac{rg^2}{\left(1-\rmi\epsilon\right)}\left[g-\frac{1}{\left(1-\rmi\epsilon\right)}-\frac{\epsilon^2}{\left(1-\rmi\epsilon\right)}\right] ,
\end{equation}

Putting all substitutions apart from~(\ref{eq:substausigma}) into
\eref{eq:recur_F}, and defining $G = r(1-\rmi\epsilon)^{-1}F$, we
obtain
\begin{equation}
  \label{eq:Gstarteqn}
  \fl \frac{G}{r} = \frac{g}{1-rg-\frac{r\sigma}{(1-\rmi\epsilon)}}
  + \frac{\sigma}{(1-rg)(1-\rmi\epsilon)-r\sigma}
  + \frac{\tau}{(1-r\gh)(1-\rmi\epsilon)-r\tau}.
\end{equation}
Though it is clear that the last two terms together are symmetric, this 
still needs to be checked for the first term.  We can rewrite it as
\begin{equation}
\fl \frac{g}{1-rg-\frac{r\sigma}{(1-\rmi\epsilon)}} = \frac{g(1-\rmi\epsilon)}{1-\rmi\epsilon -r\sigma -rg(1-\rmi\epsilon)} = \frac{g\gh(1-\rmi\epsilon)}{\gh(1-\rmi\epsilon -r\sigma) -rg\gh(1-\rmi\epsilon)} ,
\end{equation}
so that we can see the symmetry follows from \eref{eq:rel_g_ghat}.

Having verified the symmetry of the generating function $G$ we can put
in the symmetry-breaking substitution \eref{eq:substausigma}, or
rather the simpler and equivalent $\tau=0$ and
$\sigma=\rmi\epsilon-1$.  We finally get the generating function of
the required semiclassical correlation functions
\begin{equation}
  \label{eq:recur_G}
  G = \frac{r(g-1)}{1-r(g-1)}, \qquad G = \sum_{n\geq1}r^{n} D(\epsilon,n).
\end{equation}

From \eref{eq:mom_thru_D}, we can see that the $n$-th moment is simply
$(\rmi\mu)^{-n}$ times the coefficient of $(\epsilon r)^{n}$ in the
expansion of $G$.  Setting $a=\rmi\epsilon$ to cancel the factors of
$\rmi$, we just need to extract the coefficients of $(ar)^{n}$ from
$G$.  To do so we let $y = r(g-1)$ and obtain
\begin{equation}
  \label{eq:recur_G_in_y}
  G = \frac{y}{1-y} , \qquad y = \frac{G}{1+G}.
\end{equation}
Substituting $g=1+y/r$, $r=s / a$ and setting $a=0$ in
equation~\eref{gnewgenfunctioneqn}, we get the equation
\begin{equation}
  \label{eq:almost_final_eq}
  s - ys - y + 2 y^2 = 0 ,
\end{equation}
for $y$ and, through relation~\eref{eq:recur_G_in_y}, the equation
\begin{equation}
  \label{eq:final_eq}
  G^2 + (s-1)G + s = 0 ,
\end{equation}
for the generating function of the moments.  Solving this, and taking
the solution which gives the correct value of $G=0$ when $r=0$, we get
exactly the generating function \eref{rmtgenfunction} from RMT.  As
the semiclassically calculated moments match the RMT ones to all
orders, we then recover the full distribution of the delay times
\eref{rmttaudisteqn} from \cite{bfb99}.

\section{Conclusions and Outlook}

For open chaotic systems we presented a semiclassical derivation of all 
moments of the individual delay times, up to leading order in inverse channel
number.  The derivation essentially relies on the twin properties of hyperbolicity 
and ergodicity of the chaotic dynamics: the latter for the possible return to nearby 
points to create encounters and the former to allow the reconnections inside those 
encounters. This leads to a result which is in agreement with the RMT prediction
for the delay time distribution, a semicircle type law.  Notably this 
implies an upper bound on the longest delay time.  In the
derivation we relied heavily on the previous work on semiclassical
expansions, in particular \cite{heusleretal06,mulleretal07}, which contain an
implicit assumption of instant equilibration in the underlying
classical dynamics.  The influence of slower equilibration can be
explored by treating the effect of the Ehrenfest time on the
semiclassical contributions \cite{kuipersetal09,waltneretal09}.  When
the Ehrenfest time becomes much larger than the typical time
trajectories spend inside the system one recovers the classical
exponential distribution of delay times.

It is also interesting to explore the effect of moving from chaotic to
mixed phase space (with regular islands) on the distribution of the
delay times and this upper bound.  The interference effects which lead
to the current result rely on the chaotic dynamics and should be
suppressed if the chaotic part is reduced.  However, additional
effects such as periodic orbit bifurcations \cite{keatingetal07} can
also be fairly strong. The moments of the delay times and in
particular their upper bound could therefore be very sensitive
measures to explore the dynamics inside quantum dots, and possibly
used to measure the relative weights of the chaotic and regular parts of
phase space.

By considering scattering matrix correlation functions, which were then treated 
semiclassically, we were able to derive equations that implicitly define the 
generating functions of the moments.  It is worth noting here that such correlation functions
are useful for investigating other questions, such as the density of
states of chaotic Andreev billiards.

Restricting our attention to the leading order in inverse channel
number resulted in limiting the contributing diagrams to trees only.
This, in turn, allowed standard recursive tools to be used.  Looking beyond 
the leading order, the semiclassical diagrams have more complicated structures 
so that such tools can no longer be directly applied.  Thus, in our view, the remaining 
challenge of obtaining a semiclassical expansion of all moments to all orders is 
a task of significant technical difficulty.  However, it is a task of particular interest, 
not only because the subleading orders should be influenced by the symmetries 
of the system, but also because the effects of a finite number of channels is of 
much experimental relevance.  A solution to this problem would have to involve 
new combinatorial tools and we hope that a clear and general algebraic structure will 
emerge as a result of research in this direction.

\ack{The authors would like to thank Cyril Petitjean, Misha
  Polianski and Daniel Waltner for useful discussions and gratefully acknowledge the
  Alexander von Humboldt Foundation (JK) and the National Science
  Foundation under Grant No.~0604859 (GB) for funding.}

\appendix
\section{Correlation coefficients $C(\epsilon,n)$}
\label{sec:C_all_orders}

Here we briefly outline the results for the correlation functions $C(\epsilon,n)$ defined in \eref{Cetaneqn}.  As these do not involve subtracting the identity matrix from each bracket (as we did in \eref{eq:D_def}) we no longer need to change the contribution of certain diagonal pairs or subtract anything when the start channels coincide.   We can therefore simply set both $\sigma=\tau=0$ in the treatment of section~\ref{allmoments} and generate $C(\epsilon,n)$ instead.  This simplification means that $g=\gh$ because of \eref{eq:rel_g_ghat}, so that $g$ is now given implicitly by
\begin{equation} \label{ggenfunctioneqn}
\left[g-\frac{1}{\left(1-\rmi\epsilon\right)}\right]=\frac{rg^2}{\left(1-\rmi\epsilon\right)}\left[g-\frac{1}{\left(1-\rmi\epsilon\right)}-\frac{\epsilon^2}{\left(1-\rmi\epsilon\right)}\right] ,
\end{equation}
where in fact the only difference is that the $r$ on the left hand side of \eref{gnewgenfunctioneqn} has disappeared.  We can generate the first few terms in the expansion $g=\sum r^{n-1}g_{n}$ as
\begin{equation}
g_{0}=\frac{1}{\left(1-\rmi\epsilon\right)} , \qquad
g_{1}=\frac{-\epsilon^2}{\left(1-\rmi\epsilon\right)^4} , \qquad
g_{2}=\frac{\epsilon^2\left(2\epsilon^2-1\right)}{\left(1-\rmi\epsilon\right)^7} .
\end{equation}
Adding the contribution from the top node in $F$, we obtain the full generating function of $C(\epsilon,n)$, which is given by
\begin{equation}
G(\epsilon,r)=\frac{rg}{1-rg}=rg+r^2g^2+\ldots .
\end{equation}
The expansion $G(\epsilon,r)=\sum r^{n} C(\epsilon,n)$ then leads to
%
\begin{eqnarray}
C(\epsilon,1)&=g_{0}=\frac{1}{\left(1-\rmi\epsilon\right)} ,\\
C(\epsilon,2)&=\left[g_{1}+{g_{0}}^{2}\right]=\frac{1-2\rmi\epsilon-2\epsilon^2}{\left(1-\rmi\epsilon\right)^4} ,\\
C(\epsilon,3)&=\left[g_{2}+2g_{1}g_{0}+{g_{0}}^{3}\right]=\frac{1-4\rmi\epsilon-9\epsilon^2+8\rmi\epsilon^3+5\epsilon^4}{\left(1-\rmi\epsilon\right)^7} ,
\end{eqnarray}
%
which are exactly the correlation functions we can obtain by considering the diagrams explicitly as in section~\ref{firstmoments}.  Combining these results in line with expanding \eref{eq:D_def}, we recover the first three functions $D(\epsilon,n)$ calculated in section~\ref{firstmoments}.  We could also continue to generate terms to obtain the moments via \eref{newmomeqn}. 

However, the function $G(\epsilon,r)$ contains more information than just the moments, for example by setting $\epsilon=0$ we see that $g=1$. Hence
\begin{equation}
G(0,r)=\frac{r}{1-r}=r+r^2+\ldots ,
\end{equation}
which shows that
\begin{equation}
\Tr\left[S^{\dagger}\left(E\right)S\left(E\right)\right]^n=M ,
\end{equation}
and that the unitarity of the scattering matrix holds semiclassically for all powers $n$ to leading order in inverse channel number.  As another example, this function $G$ and the correlation functions it generates appear in the density of states of chaotic Andreev billiards \cite{kuipersetal09}.


\section*{References}

\begin{thebibliography}{10}

\bibitem{eisenbud48}
L.~Eisenbud
\newblock 1948
\newblock PhD thesis, Princeton

\bibitem{wigner55}
E.~P. Wigner
\newblock 1955
\newblock {\em Phys. Rev.}, {\textbf{98}} 145--147

\bibitem{aj07}
W.~O. Amrein and {\relax Ph}.~Jacquet
\newblock 2007
\newblock {\em Phys. Rev. A}, {\textbf{75}} 022106

\bibitem{smith60}
F.~T. Smith
\newblock 1960
\newblock {\em Phys. Rev.}, {\textbf{118}} 349--356

\bibitem{lv04b}
C.~H. Lewenkopf and R.~O. Vallejos
\newblock 2004
\newblock {\em Phys. Rev. E}, {\textbf{70}} 036214

\bibitem{ks08}
J.~Kuipers and M.~Sieber
\newblock 2008
\newblock {\em Phys. Rev. E}, {\textbf{77}} 046219

\bibitem{bb74}
R.~Balian and C.~Bloch
\newblock 1974
\newblock {\em Ann. Phys.}, {\textbf{85}} 514--545

\bibitem{val98}
R.~O. Vallejos, A.~M. Ozorio~de Almeida and C.~H. Lewenkopf
\newblock 1998
\newblock {\em J. Phys. A}, {\textbf{31}} 4885--4897

\bibitem{gutzwiller71}
M.~C. Gutzwiller
\newblock 1971
\newblock {\em J. Math. Phys.}, {\textbf{12}} 343--358

\bibitem{gutzwiller90}
M.~C. Gutzwiller
\newblock 1990
\newblock {\em Chaos in classical and quantum mechanics}
\newblock Springer, New York

\bibitem{berry85}
M.~V. Berry
\newblock 1985
\newblock {\em Proc. Roy. Soc. A}, {\textbf{400}} 229--251

\bibitem{ha84}
J.~H. Hannay and A.~M. Ozorio~de Almeida
\newblock 1984
\newblock {\em J. Phys. A}, {\textbf{17}} 3429--3440

\bibitem{sr01}
M.~Sieber and K.~Richter
\newblock 2001
\newblock {\em Phys. Scr.}, {\textbf{T90}} 128--133

\bibitem{mulleretal04}
S.~M\"uller, S.~Heusler, P.~Braun, F.~Haake and A.~Altland
\newblock 2004
\newblock {\em Phys. Rev. Lett.}, {\textbf{93}} 014103

\bibitem{mulleretal05}
S.~M\"uller, S.~Heusler, P.~Braun, F.~Haake and A.~Altland
\newblock 2005
\newblock {\em Phys. Rev. E}, {\textbf{72}} 046207

\bibitem{ks07b}
J.~Kuipers and M.~Sieber
\newblock 2007
\newblock {\em Nonlinearity}, {\textbf{20}} 909--926

\bibitem{bfb99}
P.~W. Brouwer, K.~M. Frahm and C.~W.~J. Beenakker
\newblock 1999
\newblock {\em Waves in Random Media}, {\textbf{9}} 91--104

\bibitem{rs02}
K.~Richter and M.~Sieber
\newblock 2002
\newblock {\em Phys. Rev. Lett.}, {\textbf{89}} 206801

\bibitem{heusleretal06}
S.~Heusler, S.~M\"uller, P.~Braun and F.~Haake
\newblock 2006
\newblock {\em Phys. Rev. Lett.}, {\textbf{96}} 066804

\bibitem{braunetal06}
P.~Braun, S.~Heusler, S.~M\"uller and F.~Haake
\newblock 2006
\newblock {\em J. Phys. A}, {\textbf{39}} L159--L165

\bibitem{mulleretal07}
S.~M\"uller, S.~Heusler, P.~Braun and F.~Haake
\newblock 2007
\newblock {\em New J. Phys.}, {\textbf{9}} 12

\bibitem{bhn08}
G.~Berkolaiko, J.~M. Harrison and M.~Novaes
\newblock 2008
\newblock {\em J. Phys. A}, {\textbf{41}} 365102

\bibitem{kuipersetal09}
J.~Kuipers, D.~Waltner, C.~Petitjean, G.~Berkolaiko and K.~Richter
\newblock 2009
\newblock {\em Phys. Rev. Lett.} in press, arXiv:0907.2660v2

\bibitem{lv04a}
C.~H. Lewenkopf and R.~O. Vallejos
\newblock 2004
\newblock {\em J. Phys. A}, {\textbf{37}} 131--136

\bibitem{bv88}
D.~Gouyou-Beauchamps and B.~Vauquelin
\newblock 1988
\newblock {\em RAIRO Inform. Th\'eor. Appl.}, {\textbf{22}} 361--388

\bibitem{sloanes}
N.~J.~A. Sloane
\newblock 2009
\newblock The on-line encyclopedia of integer sequences, published
  electronically at http://www.research.att.com/$\thicksim$njas/sequences/

\bibitem{miller75}
W.~H. Miller
\newblock 1975
\newblock {\em Adv. Chem. Phys.}, {\textbf{30}} 77--136

\bibitem{richter00}
K.~Richter
\newblock 2000
\newblock {\em Semiclassical theory of mesoscopic quantum systems}
\newblock Springer, Berlin

\bibitem{spg03}
H.~Schanz, M.~Puhlmann and T.~Geisel
\newblock 2003
\newblock {\em Phys. Rev. Lett.}, {\textbf{91}} 134101

\bibitem{bhn08b}
G.~Berkolaiko, J.~M. Harrison and M.~Novaes
\newblock 2008
\newblock Preprint, arXiv:0809.3476

\bibitem{waltneretal09}
D.~Waltner et al.
\newblock 2010
\newblock in preparation

\bibitem{keatingetal07}
J.~P. Keating, A.~M. Ozorio~de Almeida, S.~D. Prado, M.~Sieber and R.~Vallejos
\newblock 2007
\newblock {\em Prog. Theor. Phys. Suppl.}, {\textbf{166}} 10--18

\end{thebibliography}
\end{document}